\documentclass[aps,twocolumn,longbibliography,floatfix]{revtex4-2}
\usepackage{graphicx}
\usepackage{bm}

\begin{document}

\title{Negative Temperature in Spin Dynamics Simulations}

\author{Pui-Wai Ma}
\email{leo\_ma55@hotmail.com}
\affiliation{United Kingdom Atomic Energy Authority, Culham Science Centre, Abingdon,
Oxfordshire, OX14 3DB, United Kingdom}
\author{Sergei L. Dudarev}
\email{Sergei.Dudarev@ukaea.uk; sergei.dudarev@linacre.ox.ac.uk}
\affiliation{United Kingdom Atomic Energy Authority, Culham Science Centre, Abingdon,
Oxfordshire, OX14 3DB, United Kingdom}
\begin{abstract}
A simple and computationally efficient algorithm enables implementing negative temperature values in a spin dynamics simulation. The algorithm uses a Langevin spin dynamics thermostat with a negative damping parameter, enabling the thermalization of an arbitrary interacting spin system to the Gibbs energy distribution with a given negative temperature value. Canonical spin dynamics simulations at a negative temperature are as robust as conventional positive spin temperature simulations, providing a tool for quantitative dynamic studies of the physics of highly excited magnetic states. Two simulation case studies describing spin systems with antiferromagnetic and ferromagnetic ground states are explored. The phase transitions occurring in the negative temperature range do not necessarily exhibit similarities with their positive temperature counterparts. The transition temperatures and the character of spin alignment vary depending on the spatial range and strength of spin-spin interactions.
\end{abstract}
\maketitle

Negative temperature is a well established concept in statistical mechanics \cite{Goldman,Abragam,Ramsey,Klein,Landsberg}. The occurrence of the negative temperature phenomenon was first demonstrated experimentally by Purcell and Pound \cite{Purcell} for the nuclear spin sybsystem of a LiF crystal. Similar studies were later performed using other materials \cite{Oja,Hakonen1990,Hakonen1992,Hakonen1994}, achieving negative temperatures on the nano-Kelvin scale. 

Negative temperature is a general notion not restricted to a spin ensemble. Negative temperatures are encountered in the context of plasma states \cite{Krommes} and optical lattices \cite{Sorensen,Rapp}. A negative temperature value is associated with the formation of an ordered low entropy high energy state in an ensemble where the spectrum of energy eigenstates is bounded from above \cite{Goldman,Abragam,Ramsey,Klein,Landsberg}. From the thermodynamic definition of temperature, $1/T=\partial S(E)/\partial E$, where $S(E)$ is the entropy of a macroscopic system at energy $E$, negative temperature corresponds to a range of statistical configurations where entropy $S$ is a decreasing function of energy $E$. At the point where the derivative of $S$ changes sign and function $S(E)$ changes from monotonically increasing to monotonically decreasing, we find that $1/T=0$, or in other words, $T$ changes from $T=\infty^+$  to $T=\infty^-$. A system with negative temperature has a higher energy than a system at $T=\infty^+$, hence necessarily requiring that the spectrum of energy eigenstates is bounded from above. The highest attainable total energy state corresponds to $T=0^-$. Negative temperatures are not encountered in atomistic molecular dynamics simulations, since for particles moving in real space with kinetic energy $m{\bf v}^2/2$, velocities are not bounded and a higher average energy $E$ implies higher positive temperature.

If a system is at equilibrium, the probability $W_n$ of finding it in an eigenstate $\{n\}$ with energy $E_n$ is given by the Gibbs distribution
\begin{equation}
W_n={1\over {\cal Z}} \exp \left(-\frac{E_n}{k_B T}\right),
\label{eq1}
\end{equation}
where ${\cal Z}$ is the partition function ${\cal Z}=\sum _{n}\exp \left(-{E_n}/{k_B T}\right)$.
Equation (\ref{eq1}) imposes no constraint on the sign of $T$, which can be positive or negative, as long as the partition function can be computed. Monte Carlo simulations provide a convenient means for studying statistical mechanical properties of systems accessible to simulations at negative temperature, but such studies are limited \cite{Krommes}.

In a particular example of nuclear spins \cite{Purcell,Oja,Hakonen1990,Hakonen1992,Hakonen1994}, since the strength of spin-spin interactions is low, a negative temperature state can be created by applying a strong external magnetic field $\mathbf{H}_{ext}$.  In this strong magnetic field limit, the nuclear spin Hamiltonian is dominated by the Zeeman term ${\cal H}_Z=-\mathbf{I}\cdot\mathbf{H}_{ext}$, where $\mathbf{I}$ is the nuclear spin vector. In accord with Eq. (\ref{eq1}), once the direction of external field is instantaneously reversed, a metastable short-lived population of inverted spin orientations is formed, characterized by a negative temperature value.

The external field reversal technique is what has so far been used for generating negative temperature states, both experimentally and computationally \cite{Purcell,Oja,Hakonen1990,Hakonen1992,Hakonen1994,Sorensen}. The problem with this technique is that temperature remains undefined during the equilibration process, when the system dynamically relaxes from the initial highly excited state into a state of thermodynamic equilibrium at a positive temperature. Besides, the technique only applies to systems that can be manipulated by macroscopic spatially homogeneous external fields. For example, a spin system with a ferromagnetically ordered ground state cannot be transformed into a negative temperature configuration using an applied external magnetic field technique since the external field would simply change the overall orientation of the spins in response to any attempt to change the spin order.

In this study, we propose an algorithm that enables performing a controlled computationally efficient thermalization of a dynamic spin ensemble to a negative temperature. The algorithm, involving the use of a Langevin spin thermostat developed earlier \cite{Ma2011PRB,Ma2012,Ma2014}, enables performing a detailed quantitative exploration of equilibrium physics at negative temperatures. The approach is general and applies to any interacting or non-interacting dynamic statistical mechanics spin ensemble. Dynamic spin systems exhibiting anti-ferromagnetic and ferromagnetic ground states are explored, providing computational examples where the Curie-Weiss phase transition law is observed in the negative temperature range. 

Consider a system of interacting spins described by a general type Hamiltonian ${\cal H}(\mathbf{S}_1,\mathbf{S}_2,\ldots,\mathbf{S}_N)$. The corresponding Langevin equations of motion for individual spins $\mathbf{S}_i$ are \cite{Antropov,Ma2008PRB,LandauBinder,Ma2011PRB,Ma2014}
\begin{equation}
\frac{d\mathbf{S}_i}{dt} = \frac{1}{\hbar}\left[\mathbf{S}_i\times\left(\mathbf{H}_i+\mathbf{h}_i\right)-\gamma\mathbf{S}_i\times\left(\mathbf{S}_i\times\mathbf{H}_i\right)\right],
\label{eq2}
\end{equation}
where $\mathbf{H}_i=-\partial{\cal H}/\partial \mathbf{S}_i$ is the effective field acting on $i$-th spin, $\gamma$ is a damping parameter and $\mathbf{h}_i$ is a delta-correlated fluctuating field, satisfying conditions $\left<\mathbf{h}_i(t)\right>=0$ and $\left<h_{i\alpha}(t)h_{j\beta}(t')\right>=\mu\delta_{ij}\delta_{\alpha\beta}\delta(t-t')$ \cite{Ma2011PRB,Ma2012,Ma2014}. Subscripts $\alpha$ and $\beta$ denote the Cartesian components of a vector.

According to the fluctuation-dissipation theorem \cite{Chandrasekhar,Kubo,Coffey}, one can prove that in a thermal equilibrium, by identifying the energy distribution with the Gibbs distribution (Eq. \ref{eq1}), parameters $\mu$ and $\gamma$ are related through temperature $T$ as \cite{Brown,Garcia,Ma2008PRB,Ma2010PRE,Ma2012}:
\begin{equation}
\mu=2\gamma\hbar k_B T.
\label{eq3}
\end{equation}
If $T$ is negative and $\gamma$ is positive, a seemingly straightforward idea is to assign a negative value to $\mu$. However, $\mathbf{h}_i$ then becomes a complex number vector. If one plugs such a complex fluctuating field into Eq. (\ref{eq2}), this makes $\mathbf{S}_i$ complex-valued, a clearly unnecessary complication in the context of a classical spin dynamics considered here. 

An alternative, simple yet computationally efficient and robust, solution is proposed below.  We assign a negative value to the parameter $\gamma$ instead, with $\mu$ remaining positive. If we now examine the corresponding Fokker-Planck equation \cite{Ma2010PRE} corresponding to Langevin equation  (\ref{eq2}), we see that this imposes no constraint on the sign of $\gamma$ and still enables attaining a Gibbs distribution with an arbitrary chosen $T$.

In terms of microscopic fluctuating spin dynamics, we note that if $\gamma$ is negative, the dissipation term in Eq. (\ref{eq2}) pushes the spin vector $\mathbf{S}_i$ away from it being aligned with the effective field $\mathbf{H}_i$ acting on it. This necessarily injects extra energy into the spin system, in full agreement with the notion that a negative temperature corresponds to a high-energy excited thermodynamic state of a spin ensemble. We note that if the method were to be applied to a system where the spectrum of energy states is not bounded from above, the Langevin term would continue pumping energy into the system {\it ad infinitum}.

A remarkable feature of the Langevin algorithm involving a negative friction parameter $\gamma$ and negative temperature $T$ is that it is now the friction term that delivers energy to the statistical system. This implies that the fluctuation term acts as effective damping, reversing the conventional meaning of the two (fluctuation and dissipation) terms involved in a Langevin dynamics simulation \cite{Mazo,Coffey,Dudarev2008}. 

In what follows, we consider two case studies illustrating applications of the algorithm. Both cases involve an ensemble of interacting spins described by a general non-collinear Heisenberg Hamiltonian
\begin{equation}
{\cal H} = -\frac{1}{2}\sum_{i,j}J_{ij}\mathbf{S}_i\cdot\mathbf{S}_j
\label{eq4}
\end{equation}
where $\mathbf{S}_i=\mathbf{M}_i/g\mu_B$ is an atomic spin vector, $\mathbf{M}_i$ is the corresponding atomic magnetic moment, $g=2.0023$ is the electronic g-factor, $\mu_B$ is the Bohr magneton and $J_{ij}$ is the exchange coupling parameter \cite{Turek} between spins $i$ and $j$. Although there are no literature data describing negative temperature states involving atomic spin configurations, it does not appear impossible that an application of advanced experimental techniques of spin flipping using high energy laser pulses \cite{Koopmans2009natmat} could generate highy excited non-collinear negative temperature spin states in an atomic ensemble. 
We note that the Ruderman-Kittel interaction for nuclear spins \cite{Oja} adopts the same functional form for the Hamiltonian. Therefore, the following discussion is equally applicable to nuclear as well as to atomic spins. Also, since a similar Hamiltonian formalism applies also to alloys \cite{NguyenManh2008,Lavrentiev2016}, we note the general applicability of the negative temperature notion to alloy configurations. 

In the first case study, we assume that $J_{ij}\neq 0$ only for the first nearest neighbours, choosing the Heisenberg parameter and magnetic moment values of  $J_1=-69.8$ meV and $|\mathbf{M}_i|=0.89$ $\mu_B$. In the second case study, we assume $J_1=22.52$ meV, $J_2=17.99$ meV and $|\mathbf{M}_i|=2.2$ $\mu_B$. Both types of interaction are realized on a BCC lattice. The two sets of parameters selected above refer to pure chromium and pure iron that have collinear antiferromagnetic and collinear ferromagnetic ground states, respectively \cite{Lavrentiev2011}. We denote them as model 1 and model 2.

The simulation cells explored in the study involved 16,000 dynamic spins. Initial configurations were assumed to be the respective ground states, corresponding to $T=0^+$ K. Spin dynamics simulations were performed using the coupled dynamic equations of motion, Eq. \ref{eq2}, see Refs.  
\cite{Ma2011PRB,Ma2012,Ma2014}. The temperature of the system during a simulation was monitored using an explicit expression derived in Ref. \cite{Ma2010PRE} for a dynamic spin ensemble. As was noted in the derivation \cite{Ma2010PRE}, the resulting value of the dynamic spin temperature can be positive or negative.
\begin{equation}
T=\frac{\sum_i|\mathbf{S}_i\times\mathbf{H}_i|^2}{2k_B\sum_i\mathbf{S}_i\cdot\mathbf{H}_i}.
\label{eq5}
\end{equation}
Since the numerator in the fraction is positive definite, the value of $T$ computed using this formula is negative if the denominator is negative, i.e. if the direction of a spin $\mathbf{S}_i$ is on average anti-parallel to the direction of the effective field $\mathbf{H}_i$ acting on it.

\begin{figure}
\includegraphics[clip,trim=2cm 2cm 2cm 15cm,width=8cm]{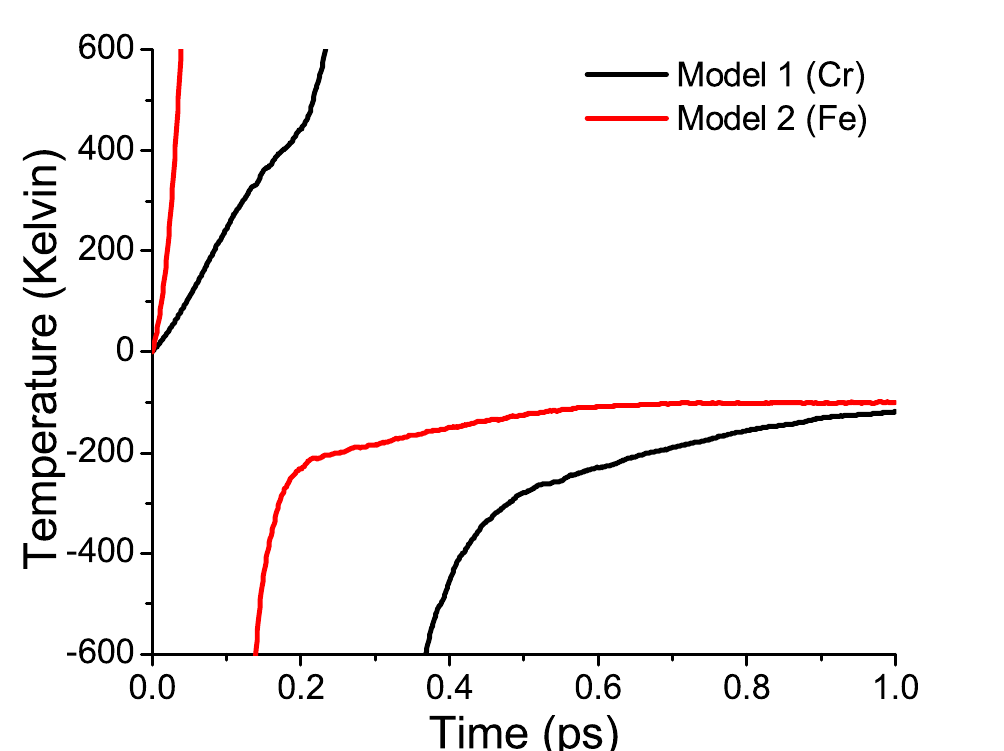}
\includegraphics[clip,trim=2cm 2cm 2cm 15cm, width=8cm]{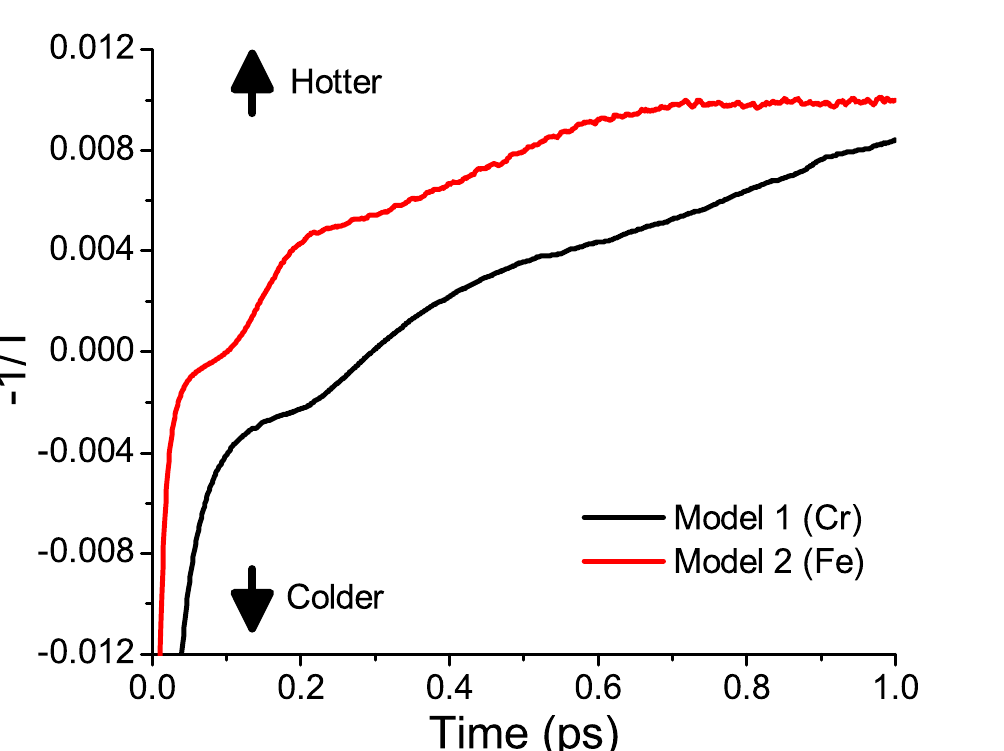}
\caption{Process of thermalization of dynamic spin ensembles with 16,000 spins from their respective ground states to $-100$ K using a Langevin thermostat with $\gamma = -0.05$. For model 1, $J_1=-69.8$ meV and $|\mathbf{M}_i|=0.89$ $\mu_B$. For model 2, $J_1=22.52$ meV, $J_2=17.99$ meV and $|\mathbf{M}_i|=2.2$ $\mu_B$. These parameters approximately describe chromium and iron with collinear antiferromagnetic and ferromagnetic ground states, respectively \cite{Lavrentiev2011,Lavrentiev2016}. (top) Evolution illustrated using the conventional Kelvin temperature scale. (bottom) Evolution illustrated using a plot involving an alternative temperature scale \cite{Goldman,Abragam,Ramsey,Klein,Landsberg}, where the vertical axis shows values of $-1/T$ in K$^{-1}$ units. }
\label{fig1}
\end{figure}

In Fig. \ref{fig1}a, we illustrate the dynamics of thermalization process for models 1 and 2 as a function of time from $0^+K$ to $-100K$, assuming $\gamma=-0.05$. Temperatures are computed on the fly using Eq. (\ref{eq5}), and are shown to approach the prescribed values at the end of a simulation. Energy is pumped into the spin system by a heat bath at $-100K$. Simulations show that temperatures first climb to $\infty^+$ and then come back from $\infty^-$.

Indeed, a spin system at a negative temperature is \textit{hotter} than the same system at any positive temperature. One can adopt an alternative temperature scale \cite{Goldman,Abragam,Ramsey,Klein,Landsberg} using a parameter $-1/T$, which offers a better description of how \textit{hot} a statistical system is. In Fig. \ref{fig1}b, we plotted temperatures during the thermalization process again, but now on a new temperature scale. Now the curves are continuous and the functional dependence is monotonic. The average energies of the spin systems are the same at $T=\infty^+$ and $\infty^-$.

\begin{figure}
\includegraphics[clip,trim=2cm 1cm 2cm 15cm,width=8cm]{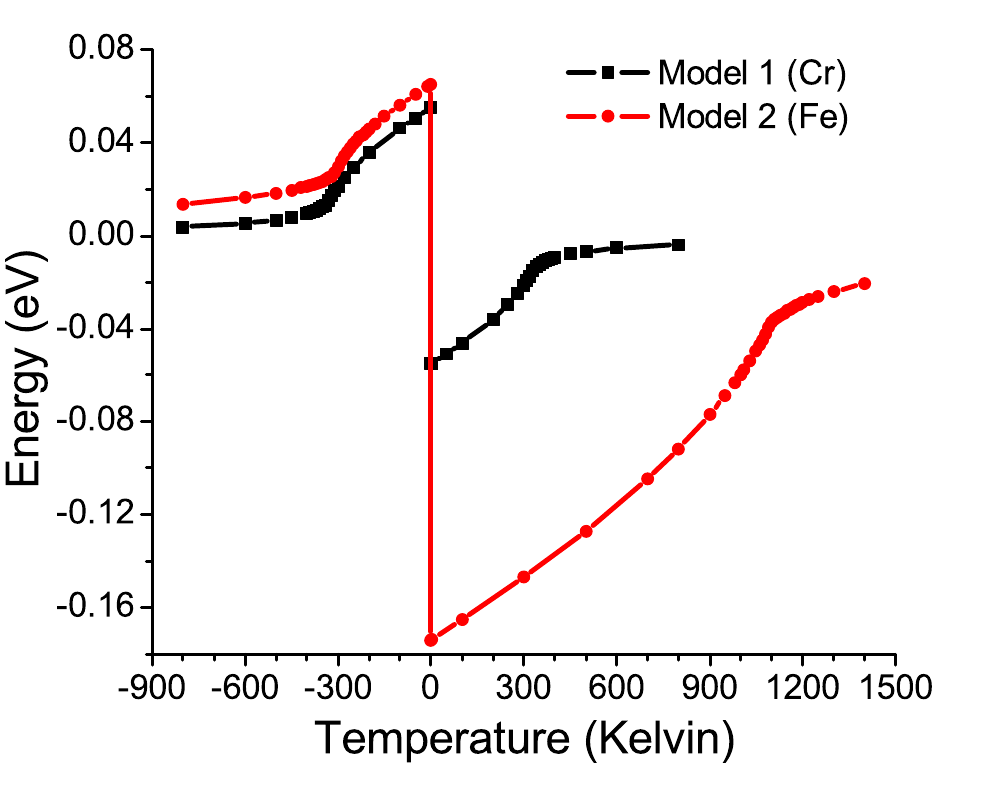}
\includegraphics[clip,trim=2cm 2cm 2cm 15cm,width=8cm]{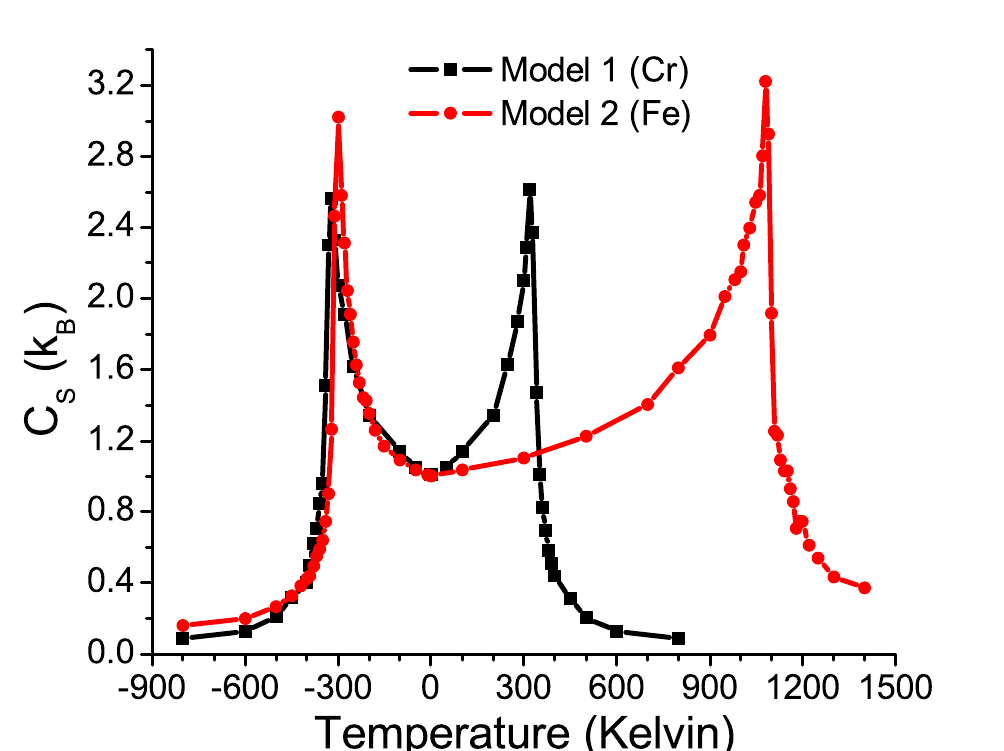}
\caption{(top) The equilibrium energy per spin as a function of temperature. (bottom) The specific heat $C_S$ is obtained by performing numerical differentiation. Peaks can be clearly observed in the negative temperature region. They correspond to negative Curie and N\'eel temperature for model 1 and 2 respectively. The asymmetric peaks for model 2 are due to the competition of $J_1$ and $J_2$ in an interacting spin system.}
\label{fig2}
\end{figure}

\begin{figure}
\includegraphics[clip,trim=1cm 0cm 1cm 10cm,height=4cm]{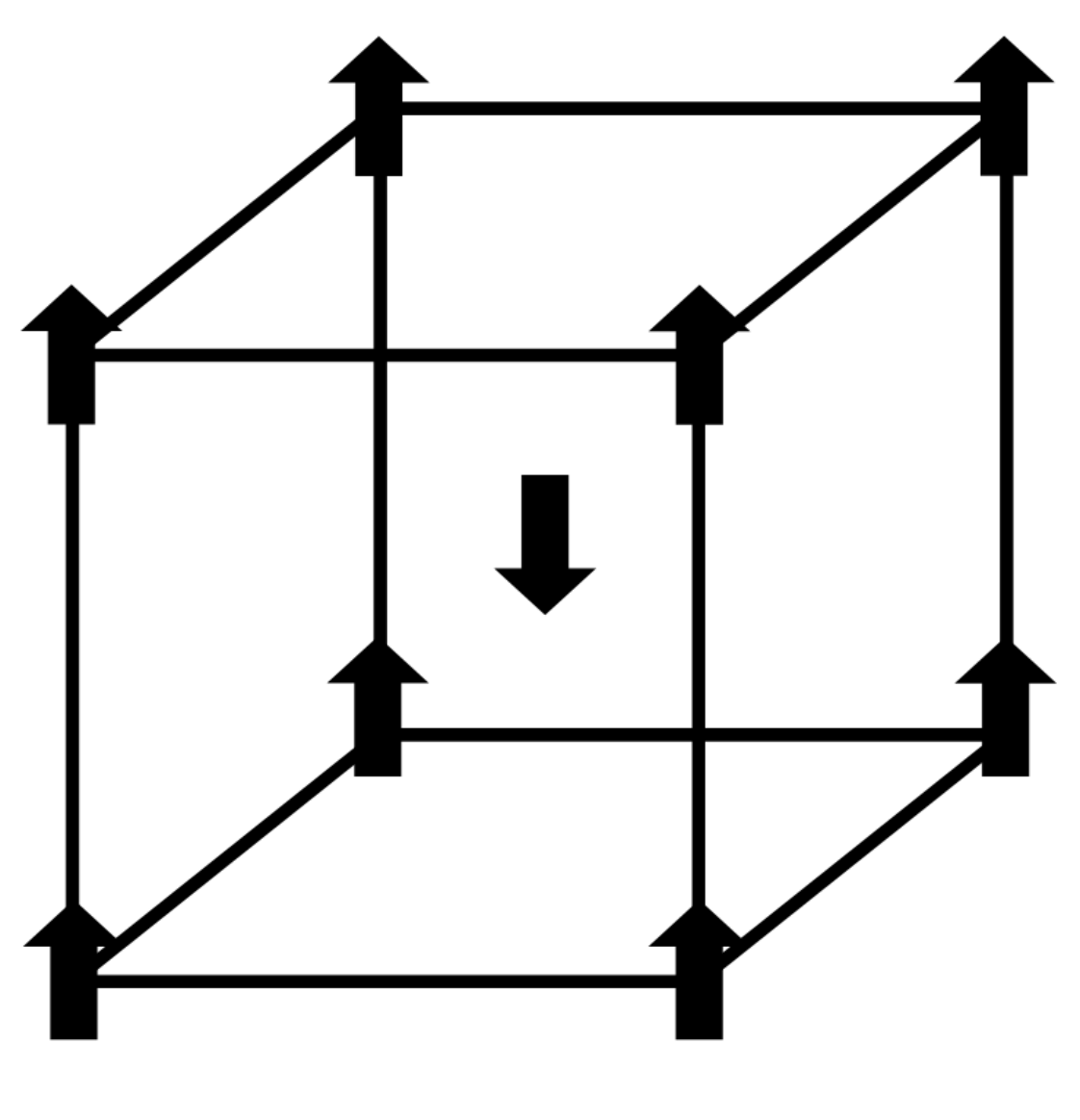}
\includegraphics[clip,trim=1cm 1cm 1cm 10cm,height=6cm]{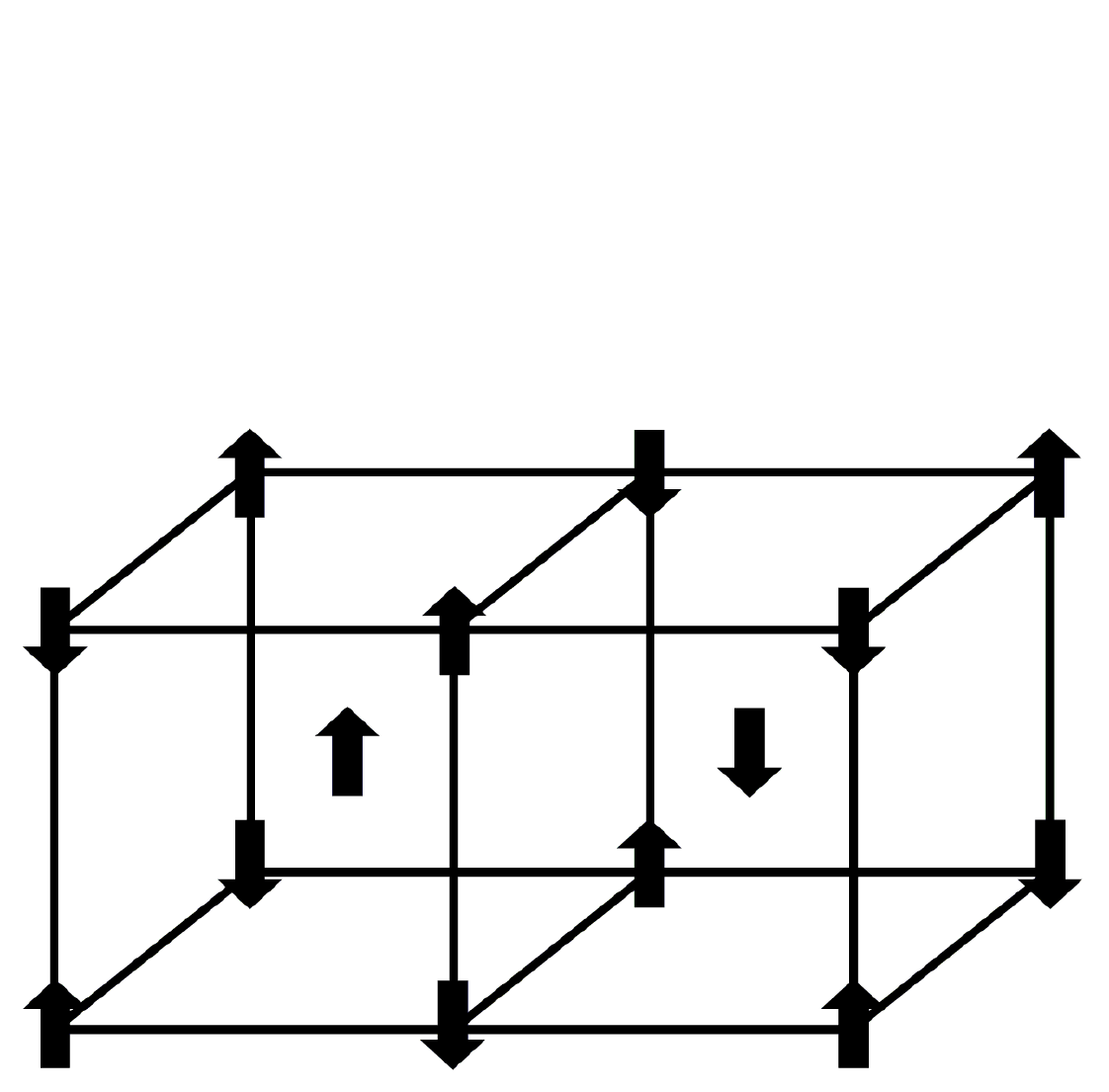}
\caption{(top) Schematic picture for antiferromagnetic type-I. (bottom) Schematic picture for antiferromagnetic type-II.}
\label{fig3}
\end{figure}

\begin{figure}
    \centering
    \includegraphics[width=8cm]{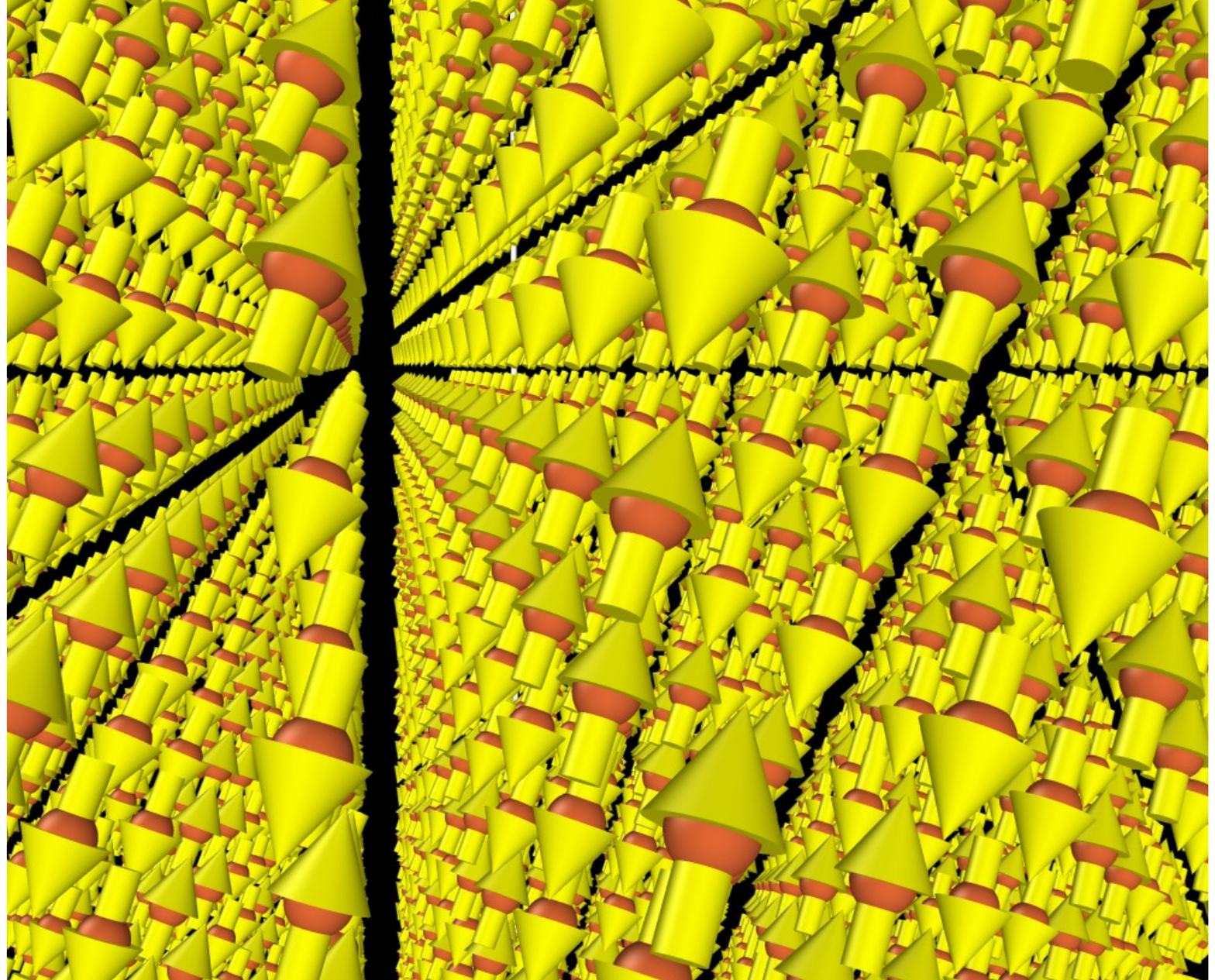}
    \caption{Atomic spins of model 2 (Fe) at $T=-1$ K.}
    \label{fig4}
\end{figure}

By thermalizing the two model systems at various temperatures to equilibrium, we find the average energy per spin as a function of temperature, plotted in Fig. \ref{fig2}a. In Fig. \ref{fig2}b we also plot the specific heat $C_S$ calculated by performing the numerical differentiation of data in Fig. \ref{fig2}a. Peaks of specific heat are observed in Fig. \ref{fig2}b in both positive and negative temperature regions. These peaks correspond to magnetic phase transitions and exhibit the characteristic shape commonly referred to as the Curie-Weiss law \cite{Desai1986}.

In the positive temperature region, the occurrence of magnetic phase transitions is well known. The peaks of the specific heat correspond to the N\'eel temperature for model 1 and to the Curie temperature for model 2, respectively. At a point where the specific heat diverges, the spin systems undergoes a phase transition from a magnetically ordered antiferromagnetic or magnetically ordered ferromagnetic state to a  paramagnetic state with no long-range spin orientational order.

In the negative temperature region, model 1 also exhibits a peak of the specific heat, illustrated in Fig. \ref{fig2}b, that is symmetric with respect to the one in the positive temperature region. The occurrence of this peak can be interpreted as follows. Since the exchange coupling $J_{ij}$ is non-zero only for the 1st nearest neighbours, if we transform $J_1\rightarrow -J_1$, the ground state of model 1 changes from an antiferromagnetic type-I (Fig. \ref{fig3}) to the ferromagnetic configuration. However, the effective field for any particular spin does not change, and as a result the absolute value of the transition temperature stays the same. Alternately, if we make the transformation $T\rightarrow -T$, while not changing $J_1$, this produces the same energy distribution (Eq. \ref{eq1}) as the reversal of the sign of $J_1$. Model 1 then exhibits a phase transition from a paramagnetic to a ferromagnetic state at a negative Curie temperature.

For model 2, the peaks are asymmetric with respect to the origin of the temperature axis. Since the exchange coupling constants $J_{ij}$ in model 2 extend to the 2nd nearest neighbour distance, by following the same logic as above for model 1, we discover that $J_1$ and $J_2$ are now in competition with each other at a negative temperature. There is no longer an obvious preferred relative arrangement of spin orientations that any pair of spins adopts at a negative temperature. 

Dynamic simulations suggest that model 2 becomes an antiferromagnetic type-II (Fig. \ref{fig3}) system when the temperature is higher than the {\it negative} N\'eel temperature. Fig. \ref{fig4} illustrates the configuration of atomic spins at $T=-1$ K. The direction of alignment of spins shown in the Figure is arbitrary, and does not correspond to the orientation of any Cartesian axis since the Hamiltonian is invariant with respect to the rotation of a spin configuration as a whole. 

The occurrence of type-II antiferromagnetic ordering can be rationalized from the energetic perspective. If model 2 adopts an antiferromagnetic type-I ordering, the energy of a magnetic moment at $T=0^-$ is $8J_1 - 6J_2=72.22$ meV. If it adopts type-II ordering, the energy per moment is $6J_2 = 107.94$ meV. Therefore, type-II ordering corresponds to higher energy, which is preferable at a negative temperature. The difference in the strength of the effective field in the positive and negative temperature regions is the reason for the lack of mirror symmetry in the peak positions in the plot for $C_S$. This result follows naturally from spin dynamics simulations but cannot be readily obtained if we consider a conventional mean-field approximation for the ferromagnetic and type-I antiferromagnetic cases.

This highlights the fact that phase transitions in the negative temperature region do not necessarily mirror their counterparts at positive temperatures. The nature of phase transitions and the transition temperatures depend on the range and strength of spin-spin interactions. 

In conclusion, in this study we develop and describe a simple and robust algorithm for the thermalization of a dynamic spin system to a negative temperature using a Langevin spin dynamic thermostat. The algorithm enables thermalizing a spin ensemble to any exact desired value of temperature, irrespectively of whether it is positive or negative. 
There is no need to introduce an external field to flip the directions of selected spins, and the energy distribution is not perturbed by the Zeeman term. 
The case studies explored above suggest that the systems with antiferromagnetic and ferromagnetic ground states exhibit phase transitions both in the positive {\it and}  negative temperature regions. Negative Curie and N\'eel temperatures were found in dynamic spin simulations. Finally, we note that although the examples considered in this study only refer to dynamic spin ensembles, the same algorithm should enable studying any dynamic statistical ensemble obeying the fluctuation-dissipation theorem.

\begin{acknowledgments}
This work has been carried out within the framework of the EUROfusion Consortium, funded by the European Union via the Euratom Research and Training Programme (Grant Agreement No 101052200 — EUROfusion) and was partially supported by the Broader Approach Phase II agreement under the PA of IFERC2-T2PA02. This work was also funded by the EPSRC Energy Programme (grant number EP/W006839/1). To obtain further information on the data and models underlying the paper please contact PublicationsManager@ukaea.uk. Views and opinions expressed are however those of the authors only and do not necessarily reflect those of the European Union or the European Commission. Neither the European Union nor the European Commission can be held responsible for them. 
\end{acknowledgments}

\bibliography{main}

\end{document}